\documentclass[a4paper,fleqn]{cas-dc}

\usepackage[authoryear]{natbib}
\usepackage{amsmath}
\usepackage{amssymb}
\usepackage{gensymb}  

\usepackage{BibDef}

\def\tsc#1{\csdef{#1}{\textsc{\lowercase{#1}}\xspace}}
\tsc{WGM}
\tsc{QE}

\begin{document}
\let\WriteBookmarks\relax
\def\floatpagepagefraction{1}
\def\textpagefraction{.001}

\shorttitle{Galaxy cores with SHARP}    

\shortauthors{Bortolas et al.}  

\title [mode = title]{Unveiling the properties of galaxy cores excavated by supermassive black hole binaries with SHARP}  



%

\author[1]{E. Bortolas}[
       orcid=0000-0001-9458-821X]
\cormark[1]
\fnmark[1]\ead{elisa.bortolas@inaf.it} 
\author[2]{E. Portaluri}
\author[3]{E. Dalla Bont\`a}
\author[1]{C. Arcidiacono}
\author[5]{P. Severgnini}
\author[6]{A. Gualandris}

\affiliation[1]{organization={INAF -- Osservatorio Astronomico di Padova, Vicolo dell’Osservatorio 5}, postcode={I-35122},
city={Padova},
country={Italy}}
\affiliation[2]{organization={INAF -- Osservatorio Astronomico d’Abruzzo, Via M. Maggini snc}, postcode={I-64100}, city={Teramo}, country={Italy}}
\affiliation[3]{organization={Dipartimento di Fisica e Astronomia ``G. Galilei”, Universit\`a di Padova, Vicolo dell’Osservatorio 3}, postcode={I-35122},
city={Padova},
country={Italy}}
\cortext[cor1]{Corresponding author}
\affiliation[4]{organization={INAF -- Osservatorio Astronomico d’Abruzzo, Via M. Maggini snc}, postcode={I-64100}, city={Teramo}, country={Italy}}
\affiliation[5]{organization={INAF - Osservatorio Astronomico di Brera, via Brera 28, 20121 Milano}, country={Italy}}

\affiliation[6]{organization={School of Mathematics and Physics, University of Surrey}, postcode={GU2 7XH}, city={Guildford}, country={UK}}

\newcommand{\msun}{\ensuremath{M_{\odot}}}

\begin{abstract}
Massive black hole  (MBH) binaries form as a result of galaxy mergers and can coalesce into a single MBH by emitting gravitational waves detectable by LISA and pulsar timing array campaigns. Although electromagnetic observations of bound MBH binaries are extremely challenging, an indirect signature of their passage is the \textit{core scouring}: a bound binary shrinks by ejecting nearby stars, creating a flat stellar density \textit{core} of the size of the binary influence radius. Through this mechanism, stars on radial orbits are preferentially ejected, resulting in a central tangential anisotropy in the velocity field of stars that can be identified via IFU observations. At present, the sample of galaxies with such properties is limited by instrument resolution to the closest giant ellipticals within the nearest $\approx100$ Mpc. The SHARP-VESPER IFU and MICADO+MORFEO instruments can work in concert to detect both these features: their unprecedented spatial resolution can allow us to detect central scourings with sizes above $\sim 500$ pc in principle up to reionization; smaller cores of $\approx 150$ pc can be detected up to $z\approx0.14$, encompassing a volume that is more than 40 times the one available at present. In addition, they can enable the search for these features in smaller galaxies, enhancing by a factor 30 the volume over which we can search for pc-size cores around $10^6-10^7\msun{}$ MBHs. The fraction of scoured galaxies, combined with their kinematic and morphological properties, carry information on the amount of merging binaries, their masses and typical environment, thus knowing this will be fundamental to complement the forthcoming gravitational wave data.
\end{abstract}





\maketitle


\section{Introduction}
\label{introduction}
Massive black hole (MBH) binaries are expected to form in large numbers as a result of the frequent galaxy-galaxy mergers that are entailed in the $\Lambda$-Cold Dark Matter model \citep{Fakhouri2010}. 
Since MBHs appear to inhabit galaxies since the onset of structure formation \citep[e.g.][]{Larson2023}, their mergers can be considered as tracers of the cosmic structure. Present and upcoming gravitational wave facilities have the capability to detect the gravitational radiation released in the coalescence of two MBHs, thus revealing their assembly across the cosmic time. In particular, the LISA instrument \citep{Amaro-Seoane2023} will be able to detect merging binaries of $10^4-10^7\msun{}$ up to $z\sim 20$ \citep{Colpi2024}, while the Einstein Telescope \citep{Abac2025} will be sensible to lower mass systems $\lesssim10^4\msun{}$ in the same redshift range. The currently operating pulsar timing array (PTA) campaigns, instead, are sensitive to the ensemble of gravitational radiation emitted by the most massive ($\gtrsim10^9\msun{}$) and relatively local ($z<1$), nearly equal mass binaries, and evidence of a gravitational wave background in the PTA band compatible with being emitted by binaries has been recently put forward \citep[e.g.][]{EPTA2024}.

Detecting {binaries} via electromagnetic radiation is extremely challenging (but see the contribution by P. Severgnini et al. about how SHARP can be transformational in such context). Although the number of observed AGN pairs has grown over the years \citep{DeRosa_et_al_2019}, such detections are limited to MBHs that are still well separated and not directly interacting;  still we lack smoking gun observations of binaries that already become a bound Keplerian pair.\footnote{Note the two MBHs can be considered to be bound roughly when the mass in stars enclosed in the binary orbit becomes comparable with the total binary mass. This occurs when the MBHs separation is of the order of the radius of  influence of the MBH binary}  An indirect  electromagnetic evidence of the (possibly past) existence of an MBH binary is the so-called core scouring process \citep{Sesana2006}. To understand its nature, we need to touch upon the mechanisms that progressively reduce the MBHs  separation \citep{Begelman1980}. 

After the two MBHs have both reached the centre of their host via  dynamical friction \citep{Chandrasekhar1943, Ostriker1999}, they form a Keplerian system. As this occurs, the slingshot ejection mechanism starts operating and quickly becomes the main driver of the binary shrinking \citep{Quinlan1996}: as stars come close to and intersect the binary orbit, they interact with it to gain positive energy \citep{Saslaw1974}, so that the binary shrinks via this mechanism, while a passing star receives a strong kick and is generally ejected from the centre of the galaxy.  In absence of other shrinking mechanisms (such as the interaction with a gaseous disk, \citealt{Souza-Lima2017}) the binary needs to eject a stellar mass of the order of its own mass to reach the separation at which gravitational waves become its main energy drainer \citep{Merritt2013}. As a direct outcome of stellar ejections, the galaxy stellar density profile is carved roughly within  the binary influence sphere, and the stellar density profile appears much flatter in this region \citep{Rantala2018, Bortolas2018tr, Dosopoulou2021}. Stars on more eccentric (radial) orbits are preferentially ejected from the centre, if compared with more circular (tangential) stellar orbits with the same energy. This feature shows up as a tangential  anisotropy bias in the velocity field of the stellar population \citep{Rantala2019}. Both the scouring and anisotropy survive past the binary coalescence \citep{Gualandris2012jan}, and can in principle be observed even several Gyr after the MBH merger has occurred{; this remains true unless intense nuclear star formation or another violent phenomenon occurs to perturb the internal dynamics of the nucleus,  thus reshuffling its internal orbital and mass configuration}.

{An additional, related mechanism that could produce a stellar core and is still related to the evolution of supermassive binaries is is the gravitational wave recoil: 
upon coalescence, asymmetric gravitational wave emission imparts  a kick velocity to the resulting MBH, which 
can  get significantly displaced from the galaxy centre. As the MBH sinks back into the centre of the galaxy, it can excavate  an even shallower and more extended core than 
scouring alone would predict, as the oscillating MBH continues 
to heat the stellar cusp while settling back to the centre 
\citep{Gualandris2008}.}

Both the core and anisotropy features have been observed,  often concurrently \citep{Rusli2013, Thomas2014}, and are virtually ubiquitous in the most massive elliptical galaxies in the local Universe.  The central lack of stars (core) is photometrically identified as the brightness profile of the host appears to be appreciably flatter in its innermost regions compared with an inward extrapolation of the more external \citet{Sersic1968} profile \citep{Savorgnan2016}. The tangential anisotropy has instead been identified within the core via integral field unit (IFU) observations and subsequent orbital reconstruction \citep{Thomas2014}. 
Since {in the scouring scenario} these features can be appreciated only within the influence sphere of the MBH, the current detections of scoured cores and tangential biases are limited to the nearest ($\sim 100$ Mpc) and to the most massive (elliptical) galaxies with the largest MBHs ($\gtrsim10^9\msun{}$), exibiting the largest influence radii (and so core sizes, of order 100 pc). These galaxies have very specific features, being the largest galaxies in the Universe: their central reservoir of cold gas is virtually empty, and so their nuclear star formation rate is negligible; in addition, they  have a stellar dynamics that is supported by dispersion, with little net rotation. Their large mass also entails the fact that they have experienced the largest amount of (major) galaxy mergers in the Universe \citep{Fakhouri2010}.

The advent of the MORFEO+MICADO \citep{MORFEO2024} and SHARP  \citep{SHARP2024} instruments, with their unprecedented spatial resolution, can substantially increase both the the number of known scoured big ellipticals, and find smaller sized cores in lower mass galaxies with smaller MBHs. This will allow for population studies on core galaxies, and it will complement the upcoming gravitional wave detections from merging MBH binaries. In the following, we outline how SHARP  on the ELT can revolutionize the observations of the dynamics on scoured cores.

\section{The study of anisotropic cores with SHARP}

\begin{figure*}
  \centering
  \includegraphics[width=0.45\textwidth]{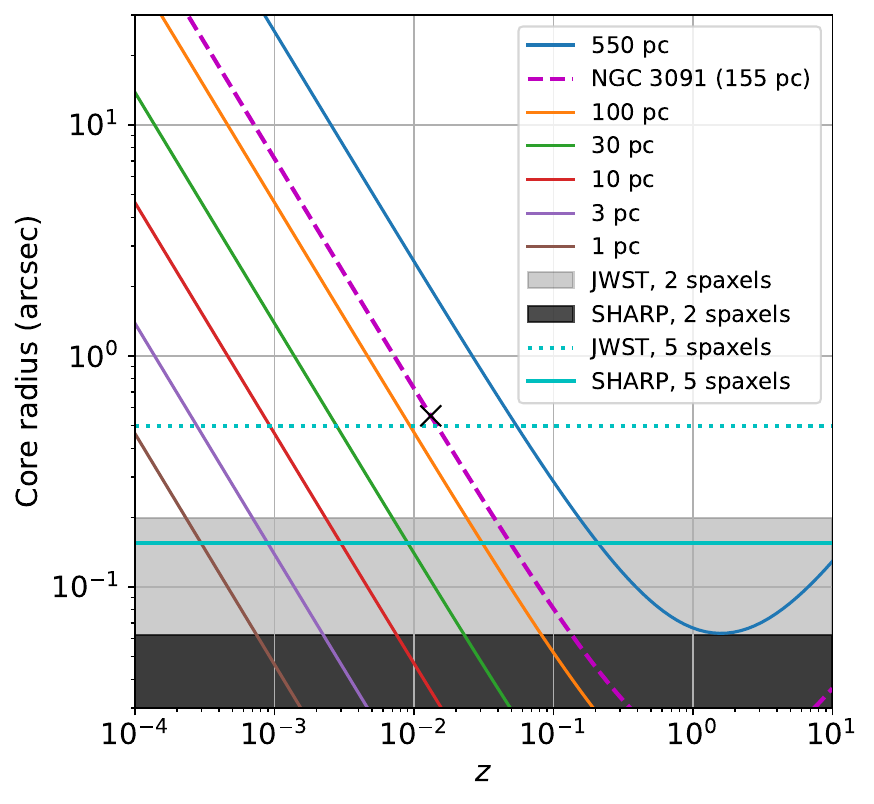}
  \includegraphics[width=0.45\textwidth]{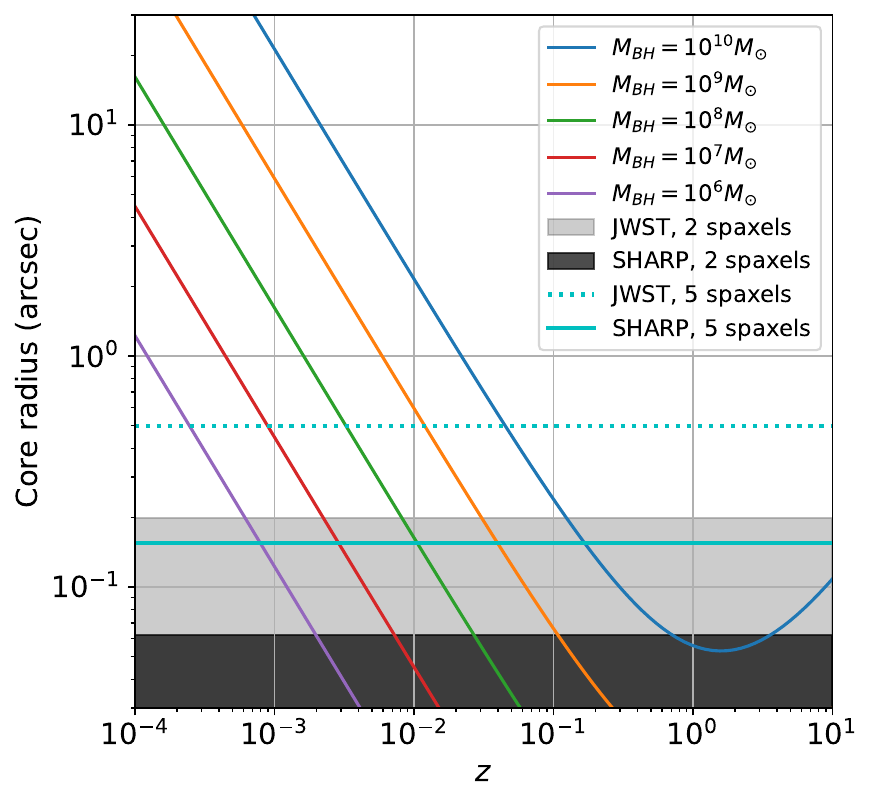}
    \caption{The plots show the angular dimensions of a core as a function of the redshift. In the left-hand  panel we quote  the size of the core in pc, while in the right-hand panel we use scaling relations in Eq.~\ref{eq:rinfl} to assign a typical size of a core for an MBH of given mass. {We also mark the angular sampling of SHARP and JWST with 2 or 5 spaxels, as shown in the legend}. Anything above these limiting areas can be resolved by either instrument. The left panel also shows the core size of a known core galaxy, NGC 3091, whose actual redshift is marked with a black "x".}\label{ffig:core_rad}
\end{figure*}

The procedure to find cores would be to first spot the dimming in the brightness profile of the galaxy through simple imaging with MORFEO and MICADO, and then confirm the nature of the core with a follow-up observation by VESPER-SHARP.
MORFEO is the ELT’s facility MCAO relay, providing uniform, wide-field correction to MICADO and the second gravity-invariant port \citep{MORFEO2024}. MORFEO will provide near-diffraction-limited performance in the
near-IR, with uniform image quality across an arcminute field of view and sufficient sky coverage to enable extragalactic science programmes at high Galactic latitude.
The SHARP spectrograph, instead, will consist of two main units:  VESPER, the multi-Integral Field Unit, and NEXUS, a  Multi-Object Spectrograph. The former unit combined with MORFEO can be used to reconstruct the structural and kinematic properties of an observed core. Each VESPER spaxel has a linear coverage of  0.031 arcseconds \citep{SHARP2024}.
{The capabilities of SHARP and MORFEO represent an improvement compared with the available IFU instruments: although IFU instruments mounted on the VLT such as MUSE NFM \citep{Weilbacher2020} and ERIS/SPIFFIER \citep{Davies2023} feature similar spaxel sizes, 
the sharpest point spread function obtainable by these instruments is several times broader 
(almost five times for the same wavelength) than MORFEO at ELT.
Furthermore, the sensitivity of MORFEO would be significantly better than that achiavable by VLT instruments.
Another possible comparison is the one between SHARP capabilities and GRAVITY+ \citep{Gravity+}, the advanced interferometer at VLTI. Although this instrument can in principle achieve fringe spacing that is only a few mas in size, it can only observe a region of  $\sim$50–60 mas simultaneously, much smaller than SHARP-VESPER field of view, that can cover $1.7''\times1.5''$ per each integral field selector, for a total field of view of nearly 31 arcsec$^2$. Although Earth rotation syntesis in GRAVITY+would allow to extend the scanned area, this would require a significant observational overhead, making it far more expensive in telescope time. SHARP-VESPER is expected to match or exceed NIRSpec IFU \citep{Jakobsen2022} in sensitivity, while delivering $\approx3\times$ finer angular resolution. For simplicity,}
in what follows we {only} compare the capabilities of SHARP with the currently operating NIRSpec IFU mounted on JWST. We assume the  \citet{Planck2014} cosmology.

A fundamental requirement to study the core of a galaxy is to be able to spatially resolve this region, which  generally extends for about the radius of influence of the MBH (binary). By using local scaling relations between the MBH mass $M_\bullet$ and the galaxy stellar velocity dispersion, the sphere of influence of the MBH
 can be shown to scale as
\begin{equation}\label{eq:rinfl}
    r_\bullet(M_\bullet) = 35 \left(\frac{M_\bullet}{10^8\msun{}}\right)^{0.56} \text{pc}
\end{equation} \citep{Merritt2009}. 
In Fig. \ref{ffig:core_rad} we show the maximum redshift at which SHARP-VESPER can resolve  the central core {with two spaxels (optimistic scenario) or five spaxels (pessimistic scenario)}, compared with the same observations if performed with JWST NIRSpec.\footnote{We use this value as accurate reconstruction of the orbital dynamics via Schwarzchild modelling within the core needs at least two to five spaxels. } 
The plots clearly show that{, in the more optimistic case, }cores of $\gtrsim500$ pc (likely found around MBHs of $\gtrsim 10^{10}\msun{}$), can be sampled with SHARP--VESPER virtually across the entire observable Universe, with the limit becoming the fact that the emission by the stellar population may get out of band and/or cosmologically dimmed and become undetectable. Let us now consider NGC 3091, a well-known core galaxy that we can use as a prototype. This galaxy is $\approx60$ Mpc from us, it hosts an MBH of $\approx 3.6\times 10^9 \msun{}$ and its core is observed to have a size of 155 pc \citep{Rusli2013}. {In the optimistic case, }the core of NGC3091 with its associated anisotropy could be observed with NIRSpec if it were located up to a luminosity distance of 172 Mpc ($z\approx0.0376$). The improved resolution of SHARP--VESPER would instead allow us to observe an analogue of this galaxy up to a luminosity distance of about 665 Mpc ($z\approx0.136$).\footnote{{In the more pessimistic scenario,the distance achiavable by NIRSpec becomes 66 Mpc ($z\approx0.0146$), while the one achiavable by SHARP-VESPER would be 227 Mpc ($z\approx0.0492$), implying an increase in the observable volume by a factor 37. }} This implies the comoving volume within which we could be able to observe it would go from 0.02 Gpc$^3$ (NIRSpec) to 0.84 Gpc$^3$ (VESPER): an increase in volume by a factor 44 implies we can in principle be able to spot many more cored galaxies with analogous properties across the cosmic epochs. Similar calculations considering a core of 300 pc result in a volume increase by a factor 62, while if we focus on the smaller cores (1-10 pc), the volume increase allowed by SHARP-VESPER compared with NIRSpec is about 33.

{We note that the observed core radii can, in some cases, exceed the predictions of Eq.~\ref{eq:rinfl} due to the additional effect of gravitational-wave recoil, which can further flatten the brightness profile beyond what is expected from binary scouring alone. By resolving the full morphological and kinematic structure of galaxy cores across a large sample, SHARP-VESPER will be uniquely positioned to statistically disentangle these two mechanisms. In particular, cores significantly larger than the influence radius of the central MBH may provide evidence that gravitational-wave recoil has occurred in a given galaxy \citep{deNicola2025}. Moreover, recoil-induced cores are expected to exhibit flatter inner profiles \citep{Khonji2024}, suggesting that measurements of the inner slope of the surface-brightness profile could help distinguish between the different excavation mechanisms. SHARP-VESPER would therefore provide a unique opportunity to identify and characterize these signatures.}

\paragraph{A focus  on NGC4889}

\begin{figure}
  \centering
  \includegraphics[width=0.45\textwidth]{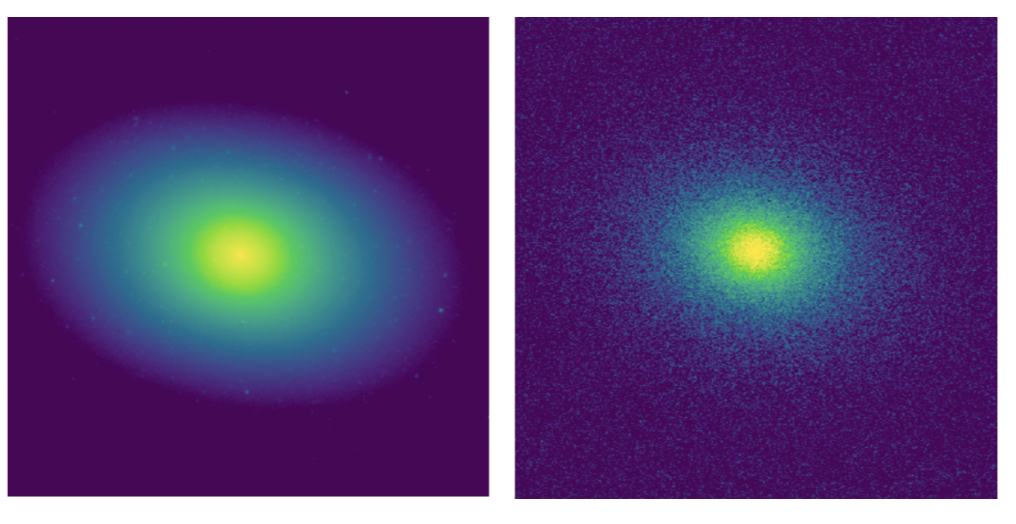}
    \caption{Left panel: {NGC 4889, located at $z=0.021$ as observed with} the WFPC2 camera through the F606W filter, with a total exposure time of 7800 s. The field of view is $21.2\times 21.2$ arcsec$^2$. Right panel: Simulated image of NGC 4889,{ moved far away in redshift, and specifically at $z$= 2.81}, obtained by using the MICADO camera in the K band under intermediate seeing conditions. The field of view of the cropped image is $1.7\times1.7$ arcsec$^2$. The colorbar is chosen to enhance the core features of the galaxy.
    }\label{fig:aetc}
\end{figure}

{

An example of observations that can be possible using MORFEO is shown in Figure~\ref{fig:aetc}, where
we considered NGC~4889, the brightest galaxy in the Coma Cluster. Previous HST-based analyses of elliptical galaxy nuclei have revealed a dichotomy between core and power-law surface brightness profiles, commonly interpreted as the result of different evolutionary pathways associated with the dissipative or stellar nature of galaxy mergers. In this context, NGC 4889 represents an optimal laboratory to investigate the dependence of core properties not only on intrinsic galaxy luminosity, but also on the cluster environment.
The galaxy is located at a redshift of z=0.021. 
Its break radius $R_b=1.58 \pm 0.05$ arcsec translates into a physical size of $R_b$= 680 pc, and is therefore well resolved by 
current and future high-resolution facilities. 
Given its high luminosity ($M_V$=-23.72 mag), NGC 4889 is particularly suited for simulations aimed at assessing the performance of next-generation instruments at higher redshifts.
In order to assess the capabilities of MORFEO+MICADO \citep{SkyCoverage} and to exploit its high spatial resolution, the galaxy was  simulated using AETC \citep{AETC} as if it were located at redshift z'=2.81, showing that its break radius at higher redshift is larger than the expected full width half maximum \citep{performances, perf_science}.

A step forward would be to properly simulate an integral field observation to explore the capability of SHARP by using, for example the python code SYNTRA (SYNthetic specTRA, \citealt{syntra}). It represents an extremely useful tool for building mock observations that can be used to recover luminosity-weighted kinematics and analyze simulations that include multiple stellar populations, kinematic cores and counter-rotating disks, and also galaxies with
both thick and thin disk components assessing the detection limits and the capability of SHARP.

}

\section{Astrophysical considerations and synergies with gravitational wave instruments}

SHARP can enhance the catalogue of known anisotropic cores in two different directions: (i) by sampling massive galaxies with very large cores up to a much higher $z$, if compared with available observations and (ii) by spotting for the first time  anisotropic, scoured cores in much smaller galaxies in the relatively local Universe. 

(i) \textit{Large cores up to high-$z$}: massive, scoured galaxies with the largest cores and MBH masses $\gtrsim10^9\msun{}$ can be observed up to cosmological distances with MORFEO and SHARP. Although such  MBHs at very high $z$ ($>4$) were thought to be very rare, recent JWST detections suggest instead that they might be much more common than previously expected \citep[see][and references therein]{Juodzbalis2024, Maiolino_2025}. \citet{Ubler2025} even observed three different AGN within a galaxy at $z\approx 5$ and with masses in the range $10^6-10^9\msun{}$, suggesting that MBH binaries or event triplets can form in the very young Universe, and thus excavate cores. Finding high-$z$ cores with the associated velocity anisotropy would be a further strong hint of the fact that MBH binaries can form, shrink and merge in a timescale shorter than the age of the Universe at that given high redshift, thus putting new constraints on the path to coalescence of MBH binaries. Yet,  it is important to mention that pristine galaxies at the dawn of structure formation may  be rich in cold gas and ongoing star formation, implying a scoured core may be quickly erased by newly formed stars as well as other violent events, such as the infall of a massive cluster near the centre. On the other hand, if an anisotropic core would be found  in the very high redshift Universe, this could be a hint of the fact that the binary is still there or has merged recently. A final relevant note on the largest cores: those are excavated by the most massive binaries, which happen to be responsible for the gravitational wave signal accessible to PTA campaigns at $z<1$. Thus collecting a demographics of the largest cores at $z<1$ would be useful to constrain which galaxies are the most likely hosts of PTA sources.

(ii) \textit{Smaller cores in the more local Universe} -- SHARP will also be able to spot smaller cores around lower mass ($10^6-10^7\msun{}$) MBHs in the nearby Universe, as described  in detail in the contribution by Dalla Bont\`a and collaborators. Although these detections would be limited to the relatively local Universe (Fig.~\ref{ffig:core_rad}), even a single observation of an anisotropic core with such a small size would be the first of its kind. In addition, the merging binaries that produce these tiny cores can be observed in the frequency band of the LISA gravitational wave detector. This implies that finding small size cores in the local Universe and performing population studies on those could provide us with hints about which galaxies are most likely to host a LISA event.

To conclude, the combination of SHARP and MICADO +MORFEO observations of galaxy cores with  LISA and PTA detections will allow us to study with unprecedented detail the population of galaxies that host or have hosted an MBH binary across the cosmic epochs.  This effort can produce new clues on the path to coalescence of merging MBHs  and their typical environments at different cosmic epochs. {A large sample scoured cores observed by these instruments would also be fundamental to test the hypothesis that massive binaries can actually excavate such large cores, if compared with the theoretically predicted amount of scoured galaxies}. This multimessenger, population-based effort will allow for a comprehensive, multimessenger  study of galaxies hosting merging MBHs.

\section*{Acknowledgments}
The SHARP team acknowledges support by Bando Ricerca Fondamentale INAF 2022,
Techno-Grant "SHARP" - 1.05.12.02.01 and Bando Ricerca Fondamentale INAF 2024,
Large-Grant "SHARP" - 1.05.24.01.01.
EP acknowledges INAF Bando di Ricerca Fondamentale INAF 2022
MiniGrant RSN1.
{During the preparation of this work the authors used ChatGPC and Claude in order to improve the text of the manuscript. After using these tools, the authors reviewed and edited the content as needed and take full responsibility for the content of the published article.}








\printcredits

\bibliographystyle{cas-model2-names}


\bibliography{bibliography} 



\end{document}